\newcounter{myctr}
\def\myitem{\refstepcounter{myctr}\bibfont\noindent\ifnum\themyctr>9\else\phantom{0}\fi\hangindent17pt\themyctr.\enskip}

\documentclass{ws-ijqi}
\usepackage{hyperref}
\usepackage[super,sort,compress]{cite}
\usepackage{xcolor}
\usepackage{appendix}
\begin{document}

\catchline{}{}{}{}{}

\title{Impossibility via  W states and feasibility via  W-like states \\
for perfect quantum teleportation}

\author{Sora Kobayashi}

\address{Graduate School of Science and Engineering, Chiba University, Chiba 263-8522, Japan\\
skobayashi@chiba-u.jp}

\author{Kei-Ichi Kondo}

\address{Department of Physics, Graduate School of Science, Chiba University, Chiba
263-8522, Japan
\\
Research and Education Center for Natural Sciences, Keio University, Kanagawa 223-8521,  Japan
\\
kondok@faculty.chiba-u.jp}

\maketitle


\begin{abstract}
We examine the two-party perfect quantum teleportation of an unknown 1-qubit state in the case of sharing various 3-qubit entangled states between a sender and a receiver: GHZ  state, W state and  W-like state. 
We give an  impossibility proof that the W state cannot be used as the sharing state to realize the perfect quantum teleportation for transmitting an arbitrary 1-qubit state, in sharp contrast with the GHZ state which is well known to realize the perfect quantum transportation.
Moreover, we give a procedure of obtaining a modified entangled state which we call the W-like state to achieve the perfect quantum transportation under a prescribed measurement basis.
\end{abstract}

\keywords{quantum information; quantum teleportation.}



\section{\label{sec:level1}Introduction}	

Since the first model of quantum teleportation was proposed by Bennet et al\cite{C},  quantum teleportation has been used as one of the most fundamental protocols in quantum communication and computation, which is now presented in standard textbooks on quantum information\cite{M}. A standard two-party quantum teleportation allows an unknown quantum state to be sent and received between two parties that share the maximum entangled state, the Bell state. That is, Alice (the sender) performs a Bell measurement on a composite system of the unknown quantum state and one of the Bell states, and informs Bob (the receiver) of the measurement result via classical communication. Then Bob receives the unknown quantum state by applying a unitary transformation to the received state of his system according to the measurement result received from Alice through the classical communication. The shared Bell states used in this procedure must be the entangled states, which is essential for Bob to be able to receive the unknown quantum state perfectly in accordance with the results of Alice's measurement. 

The entangled states generally exist in the quantum states of the $d$-qubit system $(d \geq 3)$ beyond the 2-qubit system. In particular, the entangled states of 3-qubit systems can have different entanglement configurations than the Bell states in the 2-qubit system. Indeed the GHZ and W states have different entanglements from each other\cite{G1,G2,W,V}. Since quantum teleportation is a procedure between Alice and Bob who share an entangled state, several models of quantum teleportation through the entangled states of these 3-qubit systems have been proposed\cite{VA,P,A,J,K}. However, even though perfect quantum teleportation can be constructed using entangled states with modified coefficients of the basis that make up W state\cite{P,A}, it has been suggested that the model of the two-party perfect quantum teleportation with the shared standard W state cannot be constructed. Despite being proposed for imperfect teleportation\cite{J}, it has been shown that such a model cannot be constructed if Alice wants to send an unknown 2-qubit entangled state \cite{VAA}.

In this paper, we prove the impossibility of perfect quantum teleportation sharing the W state for sending an arbitrary 1-qubit state. However, the W state can be modified (W-like state) by adjusting the expansion coefficients of the basis appearing in W state to achieve perfect quantum teleportation.

This paper is organized as follows. First, in Section \ref{sec:level2}, we provide a model of perfect quantum teleportation for sending and receiving a 1-qubit unknown pure state between two parties with a 3-qubit entangled state as a shared state. Now, we introduce the model that is followed when the GHZ state is shared. In Section \ref{sec:level3}, we prove that the model of two-party quantum teleportation provided in Section \ref{sec:level2} cannot be used for perfect quantum teleportation when the W state is a shared state, focusing on Bob's system. In Section \ref{sec:level4}, we consider an entangled state called the W-like state in which the W state is modified, so that it can be used for perfect quantum teleportation, and investigate how the corresponding Alice measurement basis should be constructed when the unitary operator appearing in Bob's system is arbitrarily set after Alice's measurement in quantum teleportation with the shared modified state, the W-like state. 
The proofs in Section \ref{sec:level3} and the method of constructing the measurement basis in Section \ref{sec:level4} are our main results. 
In Section \ref{sec:level5}, we consider an alternative method which is expected to give a simpler way to determine whether or not perfect quantum teleportation is possible using only operations performed by Alice on the state shared between Alice and Bob. Unfortunately, however, we find that this method cannot be applied to the W state to show the impossibility of perfect quantum teleportation.
Finally, in Section \ref{sec:level6}, we conclude this paper. 
We note that the results on perfect quantum teleportation obtained in this paper can also be derived from a quantum information-theoretic perspective based on entanglement entropy. A summary of this complementary analysis is presented in \ref{sec:Appendix}.


Throughout all sections in this paper, all Hilbert spaces of states are composed of two-level systems and the orthonormal basis of these Hilbert spaces is denoted by $\{|0\rangle,|1\rangle\}$.

\section{\label{sec:level2}Model of two-party quantum teleportation with shared 3-qubit entangled states}

In this section, we give an overview of two-party quantum teleportation and present an example of perfect quantum teleportation through a 3-qubit entangled state adopting the  GHZ state. 

We consider the standard two-party quantum teleportation of an unknown quantum state $|\psi\rangle_1$ by sharing the Bell state $|\beta_{00}\rangle_{23}$ between Alice and Bob: 

\begin{equation}
|\psi\rangle_1\otimes|\beta_{00}\rangle_{23} = \frac{1}{2}\sum^{1}_{m,n = 0}|\beta_{mn}\rangle_{12} \otimes \hat{U}^{(mn)}|\psi\rangle_3, \label{eq1} 
\end{equation}

where on the right-hand side we use the subscripts 1 and 2 to denote the quantum system possessed by Alice and 3 the quantum system possessed by Bob. The Bell state as the measurement basis (Bell basis) $\{|\beta_{mn}\rangle\}$ and unitary operators $\{\hat{U}^{(mn)}\}$ appearing in eq.\eqref{eq1} are given by

\begin{eqnarray}
&&|\beta_{m0}\rangle = \frac{1}{\sqrt{2}}(|00\rangle + (-1)^m|11\rangle) \ , \  
|\beta_{m1}\rangle = \frac{1}{\sqrt{2}}(|01\rangle + (-1)^m|10\rangle), \ (m = 0,1),\label{eq3} \\
&&\hat{U}^{(00)} = \hat{I} \ , \ \hat{U}^{(01)} = \hat{\sigma_x} \ , \ 
\hat{U}^{(10)} = \hat{\sigma_z} \ , \ \hat{U}^{(11)} = -i\hat{\sigma_y} . \label{eq4}
\end{eqnarray}

From eq.\eqref{eq1}, Bob obtains $\hat{U}^{(mn)}|\psi\rangle_3$ as the state of his system by receiving the information $(m,n)$ on Alice's measurement  of the basis $\{|\beta_{mn}\rangle\}$ for the composite system 1 and 2 via classical communication. This state can be converted into an initial unknown quantum state $|\psi\rangle$ at any time by Bob, no matter what the measurement result is. 

We can extend the procedure \eqref{eq1} to quantum teleportation via a 3-qubit entangled state. The perfect quantum teleportation of $|\psi\rangle_1$ sharing a 3-qubit entangled state $|\varphi\rangle_{234}$ takes the form:

\begin{eqnarray}
|\psi\rangle_1\otimes|\varphi\rangle_{234} = \sum^{1}_{\ell,m,n = 0}c_{\ell mn}|\beta_{\ell m n}\rangle_{123} \otimes \hat{U}^{(\ell mn)}|\psi\rangle_4 \ 
(c_{\ell mn} \in \mathbb{C}), \ \label{eq5} 
\end{eqnarray}

where subscripts 1, 2 and 3 of the state are the quantum system possessed by Alice and 4 the quantum system possessed by Bob, $\{|\beta_{\ell mn}\rangle\}$ represents the orthonormal basis and $\{\hat{U}^{(mn)}\}$ represents the unitary operator to be applied to Bob's system after receiving Alice's measurement results. The expansion coefficient $\{c_{\ell mn}\}$ appearing in eq.\eqref{eq5} yields the probability distribution $\{|c_{\ell mn}|^2\}$ corresponding to Alice's measurement results.

In what follows, we focus on the sharing state called the GHZ state $|\text{GHZ}\rangle$ \cite{G1,G2,VA,A}, which is a 3-qubit entangled state 
defined by

\begin{equation}
|\text{GHZ}\rangle_{234} = \frac{1}{\sqrt{2}}(|000\rangle_{234} + |111\rangle_{234} ) \label{eq6}.
\end{equation}

When the state $|\varphi\rangle$ of eq.\eqref{eq5} is equal to the GHZ state $|\text{GHZ}\rangle$, $|\varphi\rangle = |\text{GHZ}\rangle$, the perfect teleportation would be represented as

\begin{equation}
|\psi\rangle_1\otimes|\text{GHZ}\rangle_{234} = \sum^{1}_{\ell,m,n = 0}c_{\ell mn}|\beta_{\ell m n}\rangle_{123} \otimes \hat{U}^{(\ell mn)}|\psi\rangle_4 \label{eq7} .
\end{equation}

We find that the expansion coefficients $\{c_{\ell mn}\}$, the measurement basis $\{|\beta_{\ell mn}\rangle\}$ and the unitary operator $\{\hat{U}^{(\ell mn)}\}$ can be constructed by choosing, e.g.,

\begin{eqnarray}
&&c_{000} =c_{001} =c_{100} =c_{101} = \frac{1}{2} \ , \  
c_{010} =c_{011}=c_{110}=c_{111}=0,\label{eq9} \\
&&|\beta_{00m}\rangle = \frac{1}{\sqrt{2}}(|000\rangle + (-1)^m|111\rangle) \ , \ 
|\beta_{10m}\rangle = \frac{1}{\sqrt{2}}(|100\rangle + (-1)^m|011\rangle), \label{eq11} \\
&&\hat{U}^{(000)} = \hat{I} \ , \ \hat{U}^{(100)} = \hat{\sigma_x} \ , \ 
\hat{U}^{(001)} = \hat{\sigma_z} \ , \ \hat{U}^{(101)} = -i\hat{\sigma_y} \label{eq12}.
\end{eqnarray}

Substituting these expansion coefficients $\{c_{\ell mn}\}$ and unitary operators $\{\hat{U}^{(\ell mn)}\}$ into eq.\eqref{eq7}, we obtain a concrete expression:

\begin{eqnarray}
|\psi\rangle_1\otimes|\text{GHZ}\rangle_{234} =&&\frac{1}{2}|\beta_{000}\rangle_{123}\otimes|\psi\rangle_4
+ \frac{1}{2}|\beta_{100}\rangle_{123}\otimes(\hat{\sigma_x}|\psi\rangle_4) \nonumber\\
&&+\frac{1}{2}|\beta_{001}\rangle_{123}\otimes(\hat{\sigma_z}|\psi\rangle_4)
+ \frac{1}{2}|\beta_{101}\rangle_{123}\otimes(-i\hat{\sigma_y}|\psi\rangle_4) \label{eq13}.
\end{eqnarray}

We find that this is very similar to the formula eq.\eqref{eq1} for the two-party teleportation with the shared Bell state. Despite the fact that there are generally eight bases $\{|\beta_{\ell mn}\rangle\}$ spanning the 3-qubit system, Alice is able to complete teleportation by performing a basis measurement similar to eq.\eqref{eq1} that identifies just four states $\{|\beta_{000}\rangle,|\beta_{100}\rangle,|\beta_{001}\rangle,|\beta_{101}\rangle\}$. Thus, the GHZ state can be used as a sharing 3-qubit entangled state to realize perfect quantum teleportation.



\section{\label{sec:level3}Quantum teleportation via W state}

In this section, we show that in quantum teleportation via 3-qubit entangled states introduced in the previous section, the perfect quantum teleportation is impossible if the W state is chosen as the sharing state.

Any 1-qubit state $|\psi\rangle$ sent by Alice has the Bloch representation\cite{M} with two real parameters, i.e., angles $\theta,\varphi$:

\begin{equation}
|\psi\rangle_{1} = \cos \frac{\theta}{2}|0\rangle_{1} + e^{i\varphi}\sin \frac{\theta}{2}|1\rangle_{1} \ \ (\theta,\varphi \in \mathbb{R}). \label{eq14}
\end{equation}

The W state is represented in the form \cite{W}:

\begin{equation}
|\text{W}\rangle_{234} = \frac{1}{\sqrt{3}}(|001\rangle_{234} + |010\rangle_{234} +|100\rangle_{234}) \label{eq15}.
\end{equation}

As for the states of the subsystems making up this state, for example, the state of system 4 is characterized by the reduced density matrix:

\begin{equation}
\operatorname{tr}_{23} (|\text{W}\rangle_{234}\langle \text{W}|_{234}) = \frac{2}{3}|0\rangle_{4}\langle 0|_{4} +\frac{1}{3}|1\rangle_{4}\langle 1|_{4} \label{eq16}. 
\end{equation}

This form represents a mixed state, and thus the system 4 is entangled with the other systems 2 and 3. Consider the perfect quantum teleportation through this W state, i.e., $|\varphi\rangle = |\text{W}\rangle$:

\begin{equation}
|\psi\rangle_1\otimes|\text{W}\rangle_{234} = \sum^{1}_{\ell,m,n = 0}c_{\ell mn}|\beta_{\ell m n}\rangle_{123} \otimes \hat{U}^{(\ell mn)}|\psi\rangle_4  \label{eq17}. 
\end{equation}

When the left-hand side of \eqref{eq17} can be expanded into the form of the right-hand side of eq.\eqref{eq17}, we can say that the W state can be used for perfect quantum teleportation. In other words, the W state cannot be used for perfect quantum teleportation unless there exist the expansion coefficients $\{c_{\ell mn}\}$, the measurement basis $\{|\beta_{\ell mn}\rangle\}$ and the unitary operator $\{\hat{U}^{(\ell mn)}\}$ satisfying eq.\eqref{eq17}. 

In what follows, we prove that, given an arbitrary Alice's measurement basis $\{|\beta_{\ell mn}\rangle\}$, the conditions imposed on the expansion coefficients $\{c_{\ell mn}\}$ and operators$\{\hat{U}^{(\ell mn)}\}$  derived from the left-hand side of eq.\eqref{eq17} are inconsistent with the right-hand-side of eq.\eqref{eq17}, which means the impossibility of the perfect quantum teleportation sharing the W state.

By inserting the completeness relation ($\hat{I} =\sum^{1}_{\ell,m,n = 0}|\beta_{\ell mn}\rangle_{123}\langle\beta_{\ell mn}|_{123}$) of Alice's measurement basis into the left-hand side of eq.\eqref{eq17}, we obtain

\begin{eqnarray}
|\psi\rangle_1\otimes|\text{W}\rangle_{234}
=&&\Bigg(\sum^{1}_{\ell,m,n = 0}|\beta_{\ell mn}\rangle_{123}\langle\beta_{\ell mn}|_{123}\Bigg)|\psi\rangle_1\otimes|\text{W}\rangle_{234} \nonumber\\
=&&\sum^{1}_{\ell,m,n = 0}|\beta_{\ell mn}\rangle_{123}\otimes \hat{T}^{(\ell mn)}|\psi\rangle_4 \label{eq18} , 
\end{eqnarray}

where we have defined the operator $\{\hat{T}^{(\ell mn)}\}$ by

\begin{eqnarray}
\hat{T}^{(\ell mn)}|\psi\rangle_4 = {}_{123}\langle\beta_{\ell mn}|(|\psi\rangle_1\otimes|\text{W}\rangle_{234}). \label{eq17.55}
\end{eqnarray}

On the other hand, the left-hand side of eq.\eqref{eq17} is explicitly written using eq.\eqref{eq14} and eq.\eqref{eq15} as

\begin{eqnarray}
|\psi\rangle_1\otimes|\text{W}\rangle_{234} 
=&&\Bigg(\cos \frac{\theta}{2}|0\rangle_{1} + e^{i\varphi}\sin \frac{\theta}{2}|1\rangle_{1}\Bigg)
\otimes\frac{1}{\sqrt{3}}(|001\rangle_{234} + |010\rangle_{234} +|100\rangle_{234}) \nonumber\\
=&&\frac{1}{\sqrt{3}}\{(|001\rangle_{123} +|010\rangle_{123})|0\rangle_{4}\langle0|_{4} 
+ |000\rangle_{123}|1\rangle_{4}\langle0|_{4}  \nonumber\\
&&+(|101\rangle_{123}+|110\rangle)|0\rangle_{4}\langle1|_{4} 
+ |100\rangle_{123}|1\rangle_{4}\langle1|_{4}\}|\psi\rangle_{4}\label{eq17.5} .
\end{eqnarray}

It should be remarked that in the last equality of eq.\eqref{eq18} and eq.\eqref{eq17.5}  the quantum state $|\psi\rangle_{1}$ in system 1 is transformed to a state $|\psi\rangle_{4}$ in system 4 ($|\psi\rangle_{1}$→$|\psi\rangle_{4}$). By inserting eq.\eqref{eq17.5} into eq.\eqref{eq17.55}, we find that $\{\hat{T}^{(\ell mn)}\}$ takes the form:

\begin{eqnarray}
\hat{T}^{(\ell mn)} 
=&& \frac{1}{\sqrt{3}}\{({}_{123}\langle\beta_{\ell mn}|001\rangle_{123} +{}_{123}\langle\beta_{\ell mn}|010\rangle_{123})|0\rangle_{4}\langle0|_{4}\nonumber\\ 
&&+ {}_{123}\langle\beta_{\ell mn}|000\rangle_{123}|1\rangle_{4}\langle0|_{4} \nonumber\\
&&+({}_{123}\langle\beta_{\ell mn}|101\rangle_{123} +{}_{123}\langle\beta_{\ell mn}|110\rangle_{123})|0\rangle_{4}\langle1|_{4} \nonumber\\
&&+ {}_{123}\langle\beta_{\ell mn}|100\rangle_{123}|1\rangle_{4}\langle1|_{4}\}. \label{eq19}
\end{eqnarray}

By comparing eq,\eqref{eq17} and eq.\eqref{eq18}, therefore, we obtain the relationship:

\begin{eqnarray}
c_{\ell mn} = ||\hat{T}^{(\ell mn)}|\psi\rangle_4|| \ , \ \hat{U}^{(\ell mn)} = \frac{\hat{T}^{(\ell mn)}}{||\hat{T}^{(\ell mn)}|\psi\rangle_4||} \ , \ \nonumber\\  
\left(||\hat{T}^{(\ell mn)}|\psi\rangle_4|| := \sqrt{{}_4\langle \psi|\hat{T}^{(\ell mn)\dagger}\hat{T}^{(\ell mn)}|\psi\rangle_4}\right). \label{eq20}
\end{eqnarray}

The operator $\{\hat{U}^{(\ell mn)}\}$ in the above equation must be unitary and the operation performed by Bob after receiving the measurement result from Alice must not depend on the unknown state $|\psi\rangle$. Therefore, the operator $\{\hat{T}^{(\ell mn)}\}$ must 
satisfy the constraint:

\begin{eqnarray}
\hat{T}^{(\ell mn)\dagger}\hat{T}^{(\ell mn)} &&= \hat{T}^{(\ell mn)}\hat{T}^{(\ell mn)\dagger}
= \sqrt{r_{\ell mn}}\hat{I} \ \ (r_{\ell mn} \in \mathbb{R}_{> 0}). \label{eq21}
\end{eqnarray}

That is, the operator $\{\hat{T}^{(\ell mn)}\}$ must be proportional to the unitary operator. Assuming the existence of an operator $\{\hat{T}^{(\ell mn)}\}$ that satisfies this condition, we show that a contradiction occurs in eq.\eqref{eq17}.

Suppose that the operator $\{\hat{T}^{(\ell mn)}\}$ satisfying the condition \eqref{eq21} could be used to expand the state $|\psi\rangle_1\otimes|\text{W}\rangle_{234}$ as in eq.\eqref{eq18}. The reduced density matrix $\hat{\rho}_4$ for the system 4 can be expressed from the left-hand side of eq.\eqref{eq18} by using the W state \eqref{eq15} and $\operatorname{tr}_{1}(|\psi\rangle_{1}\langle\psi|_{1})=1$ as

\begin{eqnarray}
\hat{\rho}_4 =&& \operatorname{tr}_{123}(|\psi\rangle_{1}\langle\psi|_{1}\otimes|\text{W}\rangle_{234}\langle \text{W}|_{234}) \nonumber\\
=&&\operatorname{tr}_{1}(|\psi\rangle_{1}\langle\psi|_{1})\operatorname{tr}_{23}(|\text{W}\rangle_{234}\langle \text{W}|_{234}) \nonumber\\
=&&\operatorname{tr}_{23}(|\text{W}\rangle_{234}\langle \text{W}|_{234}) \nonumber\\
=&&\frac{2}{3}|0\rangle_{4}\langle 0|_{4} +\frac{1}{3}|1\rangle_{4}\langle 1|_{4}\label{eq22}. 
\end{eqnarray}

On the other hand, the right-hand side of eq.\eqref{eq18} yields another form $\hat{\rho'}_4$ of the reduced density matrix $\hat{\rho}_4$:

\begin{eqnarray}
\hat{\rho'}_4
=&& \operatorname{tr}_{123}\left\{\left(\sum^{1}_{\ell,m,n = 0}|\beta_{\ell mn}\rangle_{123}\otimes \hat{T}^{(\ell mn)}|\psi\rangle_4\right)
\left(\sum^{1}_{p,q,r = 0}|\beta_{pqr}\rangle_{123}\otimes\hat{T}^{(pqr)}|\psi\rangle_4\right)^\dagger\right\} \nonumber\\
=&&\sum^{1}_{p,q,r= 0}\hat{T}^{(pqr)}|\psi\rangle_{4}\langle\psi|_{4}\hat{T}^{(pqr)\dagger} \nonumber\\
=&&\sum^{1}_{p,q,r= 0}\hat{T}^{(pqr)}\Bigg(\cos^2\frac{\theta}{2}|0\rangle_{4}\langle0|_{4} + e^{-i\varphi}\cos\frac{\theta}{2}\sin\frac{\theta}{2}|0\rangle_{4}\langle1|_{4} \nonumber\\
&&+e^{i\varphi}\cos\frac{\theta}{2}\sin\frac{\theta}{2}|1\rangle_{4}\langle0|_{4} +\sin^2\frac{\theta}{2}|1\rangle_{4}\langle1|_{4}\Bigg)\hat{T}^{(pqr)\dagger} \label{eq23} .
\end{eqnarray}

Since $\hat{\rho_4}=\hat{\rho'}_4$, the state $\hat{\rho'}_4$ must be independent of the parameters $\theta,\varphi$. From eq.\eqref{eq23}, therefore we have the relation:

\begin{eqnarray}
\frac{\partial\hat{\rho'}_4}{\partial\theta} 
= &&\sum^{1}_{p,q,r= 0}\hat{T}^{(pqr)}(-2\sin \theta|0\rangle_{4}\langle0|_{4} + \frac{e^{-i\varphi}}{2}\cos\theta|0\rangle_{4}\langle1|_{4} \nonumber\\
&&+\frac{e^{i\varphi}}{2}\cos\theta|1\rangle_{4}\langle0|_{4} +2\sin \theta|1\rangle_{4}\langle1|_{4})\hat{T}^{(pqr)\dagger} = 0 \label{eq24}.
\end{eqnarray}

Choosing $(\theta ,\varphi) =(0,0),(0,\pi/2)(\pi/2,0)$, the relation \eqref{eq24} respectively read

\begin{eqnarray}
\frac{1}{2}\sum^{1}_{p,q,r= 0}\hat{T}^{(pqr)}(|0\rangle_{4}\langle1|_{4} +|1\rangle_{4}\langle0|_{4})\hat{T}^{(pqr)\dagger} &&= 0\label{eq25} ,\\
\frac{-i}{2}\sum^{1}_{p,q,r= 0}\hat{T}^{(pqr)}(|0\rangle_{4}\langle1|_{4} -|1\rangle_{4}\langle0|_{4})\hat{T}^{(pqr)\dagger} &&= 0\label{eq25.5} ,\\
-2\sum^{1}_{p,q,r= 0}\hat{T}^{(pqr)}(|0\rangle_{4}\langle0|_{4} -|1\rangle_{4}\langle1|_{4})\hat{T}^{(pqr)\dagger} &&= 0\label{eq26}.
\end{eqnarray}

For eq.\eqref{eq25}, eq.\eqref{eq25.5} and eq.\eqref{eq26} to hold simultaneously

\begin{eqnarray}
&&\sum^{1}_{p,q,r= 0}\hat{T}^{(pqr)}|0\rangle_{4}\langle1|_{4}\hat{T}^{(pqr)\dagger} 
= \sum^{1}_{p,q,r= 0}\hat{T}^{(pqr)}|1\rangle_{4}\langle0|_{4}\hat{T}^{(pqr)\dagger} =0\label{eq27} ,\\ 
&&\sum^{1}_{p,q,r= 0}\hat{T}^{(pqr)}|0\rangle_{4}\langle0|_{4}\hat{T}^{(pqr)\dagger} 
= \sum^{1}_{p,q,r= 0}\hat{T}^{(pqr)}|1\rangle_{4}\langle1|_{4}\hat{T}^{(pqr)\dagger} \label{eq28}.
\end{eqnarray}

In fact, if the operator $\{\hat{T}^{(\ell mn)}\}$ satisfies eq.\eqref{eq27} and eq.\eqref{eq28}, then eq.\eqref{eq23} reduces to

\begin{equation}
\hat{\rho'}_4 = \sum^{1}_{p,q,r= 0}\hat{T}^{(pqr)}|0\rangle_{4}\langle0|_{4}\hat{T}^{(pqr)\dagger}, \label{eq29}
\end{equation}

which implies that the state $\hat{\rho'}_4$ itself is indeed independent of the parameters $\theta,\varphi$. Equating $\hat{\rho}_4$ of \eqref{eq22} and $\hat{\rho'}_4$ of \eqref{eq29}, we obtain

\begin{equation}
\frac{2}{3}|0\rangle_{4}\langle 0|_{4} +\frac{1}{3}|1\rangle_{4}\langle 1|_{4}=\sum^{1}_{p,q,r= 0}\hat{T}^{(pqr)}|0\rangle_{4}\langle0|_{4}\hat{T}^{(pqr)\dagger} \label{eq30}.
\end{equation}

By applying ${}_{4}\langle 0|\cdot|0\rangle_{4}$ and ${}_{4}\langle 1|\cdot|1\rangle_{4}$ to both sides of eq.\eqref{eq30} and using eq.\eqref{eq28}, we obtain

\begin{eqnarray}
\frac{2}{3}&&= \sum^{1}_{p,q,r= 0}|{}_4\langle0|\hat{T}^{(pqr)}|0\rangle_{4}|^2 =\sum^{1}_{p,q,r= 0}|{}_4\langle0|\hat{T}^{(pqr)}|1\rangle_{4}|^2 
 \label{eq31} , \ \\
\frac{1}{3}&&= \sum^{1}_{p,q,r= 0}|{}_4\langle1|\hat{T}^{(pqr)}|0\rangle_{4}|^2 =\sum^{1}_{p,q,r= 0}|{}_4\langle1|\hat{T}^{(pqr)}|1\rangle_{4}|^2 
 \label{eq32}. \ 
\end{eqnarray}

Here the matrix $\hat{T}^{(pqr)}$ with respect to the basis $\{|0\rangle,|1\rangle\}$ has the representation:

\begin{equation}
\hat{T}^{(pqr)}=\begin{pmatrix}T^{(pqr)}_{00}&T^{(pqr)}_{01}\\T^{(pqr)}_{10}&T^{(pqr)}_{11}
\end{pmatrix} ,
\end{equation} 

where each component $T^{(pqr)}_{jk}$ is defined by $T^{(pqr)}_{jk} := \langle j|\hat{T}^{(pqr)}|k\rangle$. Therefore, the constraint eq.\eqref{eq21} is rewritten into

\begin{eqnarray}
\hat{T}^{(pqr)}\hat{T}^{(pqr)\dagger} 
=&&\begin{pmatrix}T^{(pqr)}_{00}&T^{(pqr)}_{01}\\T^{(pqr)}_{10}&T^{(pqr)}_{11}
\end{pmatrix}
\begin{pmatrix}T^{(pqr)*}_{00}&T^{(pqr)*}_{10}\\T^{(pqr)*}_{01}&T^{(pqr)*}_{11}
\end{pmatrix} \nonumber\\
=&&\begin{pmatrix}|T^{(pqr)}_{00}|^2 + |T^{(pqr)}_{01}|^2&T^{(pqr)}_{00}T^{(pqr)*}_{10} + T^{(pqr)}_{01}T^{(pqr)*}_{11}\\T^{(pqr)}_{10}T^{(pqr)*}_{00}+T^{(pqr)}_{11}T^{(pqr)*}_{01}&|T^{(pqr)}_{10}|^2 + |T^{(pqr)}_{11}|^2
\end{pmatrix} \nonumber\\
=&&\sqrt{r_{pqr}}\begin{pmatrix}1&0\\0&1
\end{pmatrix}.
\end{eqnarray} \label{eq33.5}
Therefore, we have the relation:
\begin{eqnarray}
|T^{(pqr)}_{00}|^2 + |T^{(pqr)}_{01}|^2=|T^{(pqr)}_{10}|^2 + |T^{(pqr)}_{11}|^2, \label{eq33}
\end{eqnarray}

which is further rewritten from the definition of $T^{(pqr)}_{jk}$ as 

\begin{eqnarray}
|{}_4\langle0|\hat{T}^{(pqr)}|0\rangle_{4}|^2 +|{}_4\langle0|\hat{T}^{(pqr)}|1\rangle_{4}|^2 
=|{}_4\langle1|\hat{T}^{(pqr)}|0\rangle_{4}|^2 +|{}_4\langle1|\hat{T}^{(pqr)}|1\rangle_{4}|^2. \label{eq34} 
\end{eqnarray}

Summing up eq.\eqref{eq34} over all $p,q,r = 0,1$ leads to

\begin{eqnarray}
&&\sum^{1}_{p,q,r= 0}(|{}_4\langle0|\hat{T}^{(pqr)}|0\rangle_{4}|^2 +|{}_4\langle0|\hat{T}^{(pqr)}|1\rangle_{4}|^2) \nonumber\\
= &&\sum^{1}_{p,q,r= 0}(|{}_4\langle1|\hat{T}^{(pqr)}|0\rangle_{4}|^2 +|{}_4\langle1|\hat{T}^{(pqr)}|1\rangle_{4}|^2). \label{eq36}
\end{eqnarray}

Therefore, substituting eq.\eqref{eq31} and eq.\eqref{eq32} into both sides of eq.\eqref{eq36} shows that the equality in eq.\eqref{eq36} does not hold. Therefore, the operator $\{\hat{T}^{(\ell mn)}\}$ that satisfies eq.\eqref{eq21} does not exist and hence cannot be used to expand the state $|\psi\rangle_1\otimes|\text{W}\rangle_{234}$ as in eq.\eqref{eq18}. Thus, irrespective of any Alice's measurement basis $\{|\beta_{\ell m n}\rangle\}$, we cannot prepare expansion coefficients $\{c_{\ell mn}\}$ and unitary operators $\{\hat{U}^{(\ell mn)}\}$ that satisfy eq.\eqref{eq17}, which proves that the W state cannot be used for perfect quantum teleportation as a sharing state.



\section{\label{sec:level4}W-like state suitable for the perfect quantum teleportation}

In the previous section, we have proved that 
the two-party quantum teleportation cannot be perfect if the sharing state is the standard W state \eqref{eq15}.
In this proof, we used the fact that the values (2/3,1/3) in eq.\eqref{eq31} and eq.\eqref{eq32} derived from the W state \eqref{eq15} do not allow for eq.\eqref{eq36} to hold. By adopting the same basis $(|001\rangle,|010\rangle,|100\rangle)$ as that of the W state \eqref{eq15} and changing the coefficients of the basis, the W state \eqref{eq15} is modified to be used for perfect quantum teleportation, which we call the \textit{W-like state}. Specifically, when the coefficients of the basis constructing the W state are changed so that the state of Bob is the maximum mixed state $\hat{I}/2$, the corresponding values in eq.\eqref{eq31} and eq.\eqref{eq32} become 1/2 and eq.\eqref{eq36} holds. 
In this section, we actually show the existence of such W-like states and how to construct Alice's measurement basis.

First, we construct the W-like state with the coefficients modified from the W state \eqref{eq15} in the form:

\begin{eqnarray}
|\text{W-like}\rangle_{234} = x|001\rangle_{234} + y|010\rangle_{234} + z|100\rangle_{234}  \nonumber\\ 
 (x,y,z \in \mathbb{C} ;  |x|^2 +|y|^2 +|z|^2 = 1), \label{eq39}
\end{eqnarray}

where the condition $|x|^2 +|y|^2 +|z|^2 = 1$ is required from the normalization of the W-like state. 
We proceed to find the coefficients $x,y,z$ for which the state of Bob (system 4) in this state becomes the maximum mixed state $\hat{I}/2$. The state of system 4 in the W-like state eq.\eqref{eq39} is calculated as

\begin{eqnarray}
\operatorname{tr}_{23} (|\text{W-like}\rangle_{234}\langle \text{W-like}|_{234})
= (|y|^2 +|z|^2)|0\rangle_{4}\langle 0|_{4} +|x|^2|1\rangle_{4}\langle 1|_{4}. \label{eq40} 
\end{eqnarray}

The coefficients $x,y,z$ to yield the maximum mixed state $(|0\rangle_{4}\langle 0|_{4} + |1\rangle_{4}\langle 1|_{4})/2=\hat{I}/2$ are obtained by the general solution for the simultaneous equations: $|x|^2=1/2$, $|y|^2 +|z|^2=1/2$ and $|x|^2 +|y|^2 +|z|^2 = 1$ as

\begin{eqnarray}
x = \frac{1}{\sqrt{2}}e^{i\theta} \ , \ y = \frac{1}{\sqrt{2}}e^{i\varphi}\cos\gamma \ , \ z = \frac{1}{\sqrt{2}}e^{i\omega}\sin\gamma \ 
(\theta,\varphi,\omega,\gamma \in \mathbb{R}). \label{eq41}
\end{eqnarray}

Substituting this solution into the right-hand side of eq.\eqref{eq39} and organizing it, the W-like state is obtained:

\begin{eqnarray}
|\text{W-like}\rangle_{234} =&&\frac{1}{\sqrt{2}}e^{i\theta} (|001\rangle_{234} + e^{i(\varphi-\theta)}\cos\gamma|010\rangle_{234} \nonumber\\
&&+ e^{i(\omega-\theta)}\sin\gamma|100\rangle_{234}). \label{eq42}
\end{eqnarray}

If we ignore the total phase: $\theta = 0$, the general W-like state suitable for the perfect quantum teleportation has a simple expression:

\begin{eqnarray}
|\text{W-like}\rangle_{234} =\frac{1}{\sqrt{2}}(|001\rangle_{234} + e^{i\varphi}\cos\gamma|010\rangle_{234} + e^{i\omega}\sin\gamma|100\rangle_{234}). \label{eq43}
\end{eqnarray}

Next, we try to use the W-like state \eqref{eq43} to realize the formula for the perfect quantum teleportation:

\begin{equation}
|\psi\rangle_1\otimes|\text{W-like}\rangle_{234} = \sum^{1}_{\ell,m,n = 0}c_{\ell mn}|\beta_{\ell m n}\rangle_{123} \otimes \hat{U}^{(\ell mn)}|\psi\rangle_4.  \label{eq44} 
\end{equation}

If we can specifically set up the expansion coefficients $\{c_{\ell mn}\}$, the unitary operators $\{\hat{U}^{(\ell mn)}\}$, and Alice's measurement basis $\{|\beta_{\ell mn}\rangle\}$ so that the equality is satisfied in eq.\eqref{eq44}, then the W-like state \eqref{eq43} certainly is an entangled state that can be used for the perfect quantum teleportation.  
In fact, if we use the following four orthonormal bases

\begin{eqnarray}
|\beta_{000}\rangle_{123}=&&\frac{1}{\sqrt{2}}(e^{i\varphi}\cos\gamma|001\rangle_{123}
+e^{i\omega}\sin\gamma|010\rangle_{123}+|100\rangle_{123}),\label{eq45} \\
|\beta_{001}\rangle_{123}=&&\frac{1}{\sqrt{2}}(e^{i\varphi}\cos\gamma|001\rangle_{123}
+e^{i\omega}\sin\gamma|010\rangle_{123}-|100\rangle_{123}),\label{eq46} \\
|\beta_{010}\rangle_{123}=&&\frac{1}{\sqrt{2}}(e^{i\varphi}\cos\gamma|101\rangle_{123}+e^{i\omega}\sin\gamma|110\rangle_{123}
+|000\rangle_{123}),\label{eq47} \\
|\beta_{011}\rangle_{123}=&&\frac{1}{\sqrt{2}}(e^{i\varphi}\cos\gamma|101\rangle_{123}+e^{i\omega}\sin\gamma|110\rangle_{123}
-|000\rangle_{123}),\label{eq48} 
\end{eqnarray}

the right-hand side of eq.\eqref{eq44} reads

\begin{eqnarray}
|\psi\rangle_1\otimes|\text{W-like}\rangle_{234}
=&&\frac{1}{2}|\beta_{000}\rangle_{123}\otimes|\psi\rangle_4 +\frac{1}{2}|\beta_{001}\rangle_{123}\otimes\hat{\sigma}_z|\psi\rangle_4 \nonumber\\
&&+\frac{1}{2}|\beta_{010}\rangle_{123}\otimes\hat{\sigma}_x|\psi\rangle_4 +\frac{1}{2}|\beta_{011}\rangle_{123}\otimes\hat{\sigma}_y|\psi\rangle_4\label{eq49}.
\end{eqnarray}

Here we have chosen the expansion coefficients $c_{000} = c_{001} = c_{010} = c_{011} = 1/2, c_{100} = c_{101} = c_{110} = c_{111} = 0$, the unitary operators $\hat{U}^{(000)} = \hat{I }, \hat{U}^{(001)} = \hat{\sigma}_z, \hat{U}^{(010)} = \hat{\sigma}_x, \hat{U}^{(011)} = \hat{\sigma}_y$ and Alice's measurement basis eq.\eqref{eq45},eq.\eqref{eq46},eq.\eqref{eq47}, and eq.\eqref{eq48}.  
Then the equality of perfect quantum teleportation eq.\eqref{eq44} is established. 
Thus, it is confirmed that the W-like state \eqref{eq43}  is an example for the entangled state that can be used for perfect quantum teleportation. 
The other examples for the perfect quantum teleportation with modified W states have already been proposed \cite{P,A}.

Here we show that the W-class state proposed in the previous work\cite{P,A}:
\begin{align}
|\text{W}_n\rangle_{234}=\frac{1}{\sqrt{2 + 2n}}(\sqrt{n + 1}e^{i\delta}|001\rangle_{234} + \sqrt{n}e^{ip}|010\rangle_{234} + |100\rangle_{234}) \nonumber\\ (n \in \mathbb{R}_{\ge 0}, \ p , \delta \in \mathbb{R}) 
\label{eq*12.5}
\end{align}
is contained in our W-like state up to the phase identification.
Indeed, under the change of variables from $n \in \mathbb{R}_{\ge 0}$ to $\gamma \in \left(0,\frac{\pi}{2}\right]$ defined by
\begin{align}
\cos \gamma =  \sqrt{\frac{n}{n+1}} \ , \ \sin \gamma =\sqrt{\frac{1}{n+1}} \ \ \left(\gamma \in \left(0,\frac{\pi}{2}\right]\right),
\end{align}
the W-class state \eqref{eq*12.5} is cast into the form:
\begin{align}
|\text{W}_n\rangle_{234} =&\frac{1}{\sqrt{2}}\left(e^{i\delta}|001\rangle_{234} + \sqrt{\frac{n}{n+1}}e^{ip}|010\rangle_{234}+\sqrt{\frac{1}{n+1}}|100\rangle_{234}\right) \nonumber\\ 
=&\frac{1}{\sqrt{2}}\left(e^{i\delta}|001\rangle_{234} + e^{ip}\cos \gamma |010\rangle_{234}+\sin \gamma|100\rangle_{234}\right).
\end{align}
By multiplying both sides by $e^{-i\delta}$, we finally obtain 
\begin{align}
e^{-i\delta}|\text{W}_n\rangle_{234} = \frac{1}{\sqrt{2}}(|001\rangle_{234} + e^{i(p - \delta)}\cos\gamma|010\rangle_{234} + e^{-i\delta}\sin\gamma|100\rangle_{234}). \label{eq*12}
\end{align}
We see that the right-hand side of (\ref{eq*12}) has the same form as the W-like state \eqref{eq43}.
Moreover, the W-class state \eqref{eq*12} in the limit $n \to \infty (\gamma \to 0)$
\begin{align}
e^{-i\delta}|\text{W}_n\rangle_{234}|_{n \to \infty} = \frac{1}{\sqrt{2}}(|001\rangle_{234} + e^{i(p - \delta)}|010\rangle_{234})
\end{align}
is also contained in the W-like state \eqref{eq43} as the $\gamma = 0$ case. Therefore, our W-like state contains more states than the W-class (\ref{eq*12.5}) described above.


In many of the preceding teleportation models, the unitary transformations $\{\hat{U}^{(\ell mn)}\}$ that appear in Bob's states are all fixed to be the identity operator $\hat{I}$ and the Pauli operators $\hat{\sigma}_x,\hat{\sigma}_y,\hat{\sigma}_z$ (up to complex multiples of absolute value 1). 
In principle, it is possible to investigate how the expansion coefficients $\{c_{\ell mn}\}$ and Alice's measurement bases $\{|\beta_{\ell mn}\rangle\}$ on the right-hand side of eq.\eqref{eq44} change when we choose the other unitary transformations that appear in Bob's state.
However, if all unitary transformations to be prepared are set free, it is quite complicated to determine all expansion coefficients and Alice's measurement basis deductively from them. 

The operator as a $2\times2$ complex matrix acting on a 1-qubit state can be represented by a linear combination of the identity $\hat{I}$ and the three Pauli matrices $(\hat{\sigma}_x,\hat{\sigma}_y,\hat{\sigma}_z)$. Therefore, it is enough to prepare four kinds of independent unitary operator $\{\hat{U}^{(mn)}\}(m,n = 0,1)$ to specify Bob's state, which enables us to eliminate the redundant index $\ell$ from $\{\hat{U}^{(\ell mn)}\}$. This fact does not depend on the number $d$ of the shared $d$-qubit state, as was already shown in the GHZ state \eqref{eq13}. In view of these, without loss of generality we can set up the unitary transformation which is given by the identity and the Pauli operators acting as unitary operators  to be applied to Bob's system in what follows.
 
We denote the identity and the Pauli operators in the unified way by the symbol $\{\hat{\sigma}^{(mn)}\}$: $\hat{\sigma}^{(00)} = \hat{I},\hat{\sigma}^{(01)} =\hat{\sigma}_x,\hat{\sigma}^{(10)} = \hat{\sigma}_y,\hat{\sigma}^{(11)} = \hat{\sigma}_z$.
Then the unitary operator $\{\hat{U}^{(mn)}\}$ that appears in Bob's system is given by $\{\hat{\sigma}^{(mn)}\}$ up to an arbitrary unitary 
operator $\hat{S}$:

\begin{equation}
\hat{U}^{(mn)} = \hat{\sigma}^{(mn)}\hat{S} .\label{eq50} 
\end{equation}

Using the fact that the unitary operators $\{\hat{\sigma}^{(mn)}\}$ constitute orthonormal bases (in the Hilbert-Schmidt inner product space) of operators acting on any two-dimensional complex Hilbert space, the operator $\{\hat{\sigma}^{(mn)}\}$ is found to have the property:

\begin{equation}
\operatorname{tr}(\hat{U}^{(pq)\dagger}\hat{U}^{(mn)}) = \operatorname{tr}(\hat{\sigma}^{(pq)\dagger}\hat{\sigma}^{(mn)}) = 2\delta_{pm}\delta_{qn} .\label{eq51} 
\end{equation}

By taking into account of the dimension of the operator space, we find that there is no operator orthogonal to all of these.

The perfect quantum teleportation using the W-like state \eqref{eq43} with the operators $\{\hat{U}^{(mn)}\}$ specified by eq.\eqref{eq50} in Bob's state, obeys

\begin{equation}
|\psi\rangle_1\otimes|\text{W-like}\rangle_{234} = \sum^{1}_{m,n = 0}c_{mn}|\beta_{mn}\rangle_{123} \otimes \hat{U}^{(mn)}|\psi\rangle_4, \label{eq52}
\end{equation} 

In what follows we show that the expansion coefficients $\{c_{mn}\}$ and Alice's measurement basis $\{|\beta_{mn}\rangle\}$ are determined  such that this equality holds. 
By acting  ${}_4\langle\psi|\hat{U}^{(pq)\dagger}$ on both sides of eq.\eqref{eq52}, we have

\begin{eqnarray}
|\psi\rangle_1\otimes{}_4\langle\psi|\hat{U}^{(pq)\dagger}|\text{W-like}\rangle_{234} 
= \sum^{1}_{m,n = 0}c_{mn}|\beta_{mn}\rangle_{123} \otimes {}_4\langle\psi|\hat{U}^{(pq)\dagger}\hat{U}^{(mn)}|\psi\rangle_4 . \label{eq53} 
\end{eqnarray}

Since $|\psi\rangle$ is an arbitrary unknown quantum state, this equation holds for any of $|\psi\rangle = |0\rangle, |1\rangle$. The sum $\sum_{\psi = 0}^1$ is taken for both sides of this equation to obtain

\begin{eqnarray}
&&\sum_{\psi = 0}^1|\psi\rangle_1\otimes{}_4\langle\psi|\hat{U}^{(pq)\dagger}|\text{W-like}\rangle_{234} \nonumber\\
=&& \sum^{1}_{m,n = 0}c_{mn}|\beta_{mn}\rangle_{123}\sum_{\psi = 0}^1{}_4\langle\psi|\hat{U}^{(pq)\dagger}\hat{U}^{(mn)}|\psi\rangle_4 . \label{eq54} 
\end{eqnarray}

By using the relation \eqref{eq51}, i.e.,

\begin{equation}
\sum_{\psi = 0}^1{}_4\langle\psi|\hat{U}^{(pq)\dagger}\hat{U}^{(mn)}|\psi\rangle_4 \equiv \operatorname{tr}(\hat{U}^{(pq)\dagger}\hat{U}^{(mn)}) =2\delta_{pm}\delta_{qn} . \label{eq55}
\end{equation}

in the right-hand side of eq.\eqref{eq54}, we obtain

\begin{eqnarray}
\sum_{\psi = 0}^1|\psi\rangle_1\otimes{}_4\langle\psi|\hat{U}^{(pq)\dagger}|\text{W-like}\rangle_{234} 
=&& \sum^{1}_{m,n = 0}2\delta_{pm}\delta_{qn}c_{mn}|\beta_{mn}\rangle_{123} \nonumber\\
=&&2c_{pq}|\beta_{pq}\rangle_{123}. \label{eq56}
\end{eqnarray}

Thus, from eq.\eqref{eq54}, the product of the expansion coefficient $\{c_{mn}\}$ and Alice's measurement basis $\{|\beta_{mn}\rangle\}$ is determined from the unitary operator $\{\hat{U}^{(mn)}\}$ as 

\begin{equation}
c_{pq}|\beta_{pq}\rangle_{123} = \frac{1}{2}\sum_{\psi = 0}^1|\psi\rangle_1\otimes{}_4\langle\psi|\hat{U}^{(pq)\dagger}|\text{W-like}\rangle_{234}. \label{eq57} 
\end{equation}

Now we show that the states $\{c_{pq}|\beta_{pq}\rangle\}$ determined by this expression form an orthonormal system.
First, we observe that the inner product takes the form:

\begin{eqnarray}
{}_{123}\langle\beta_{mn}|c_{mn}^*c_{pq}|\beta_{pq}\rangle_{123}
 =&& \left(\frac{1}{2}\sum_{\tilde{\psi} = 0}^1|\tilde{\psi}\rangle_1\otimes{}_4\langle\tilde{\psi}|\hat{U}^{(mn)\dagger}|\text{W-like}\rangle_{234}\right)^\dagger \nonumber\\
&&\times\left(\frac{1}{2}\sum_{\psi = 0}^1|\psi\rangle_1\otimes{}_4\langle\psi|\hat{U}^{(pq)\dagger}|\text{W-like}\rangle_{234}\right) \nonumber\\
=&&\frac{1}{4}\sum_{\psi,\tilde{\psi} = 0}^1\delta_{\psi\tilde{\psi}} \ {}_{234}\langle \text{W-like}|\hat{U}^{(mn)}|\tilde{\psi}\rangle_4{}_4\langle\psi|\hat{U}^{(pq)\dagger}|\text{W-like}\rangle_{234}\nonumber\\
=&&\frac{1}{4}{}_{234}\langle \text{W-like}|\hat{U}^{(mn)}\hat{U}^{(pq)\dagger}|\text{W-like}\rangle_{234}\nonumber\\
=&&\frac{1}{4}{}_{234}\langle \text{W-like}|\hat{\sigma}^{(mn)}\hat{\sigma}^{(pq)\dagger}|\text{W-like}\rangle_{234} ,\label{eq58} 
\end{eqnarray}

where we have used eq.\eqref{eq50} in the last step. Next, we observe that the states $\{\hat{\sigma}^{(pq)\dagger}|\text{W-like}\rangle_{234}\}$ form an orthonormal system:

\begin{equation}
{}_{234}\langle \text{W-like}|\hat{\sigma}^{(mn)}\hat{\sigma}^{(pq)\dagger}|\text{W-like}\rangle_{234} = \delta_{mp}\delta_{nq}. \label{eq63} 
\end{equation}

This fact can be checked by explicitly writing the states:

\begin{eqnarray}
\hat{\sigma}^{(00)\dagger}|\text{W-like}\rangle_{234} =&& \frac{1}{\sqrt{2}}(|001\rangle_{234} + e^{i\varphi}\cos\gamma|010\rangle_{234} 
+ e^{i\omega}\sin\gamma|100\rangle_{234}), \label{eq59} \\
\hat{\sigma}^{(01)\dagger}|\text{W-like}\rangle_{234} =&& \frac{1}{\sqrt{2}}(|000\rangle_{234} + e^{i\varphi}\cos\gamma|011\rangle_{234} 
+ e^{i\omega}\sin\gamma|101\rangle_{234}), \label{eq60} \\
\hat{\sigma}^{(10)\dagger}|\text{W-like}\rangle_{234} =&& \frac{i}{\sqrt{2}}(-|000\rangle_{234} + e^{i\varphi}\cos\gamma|011\rangle_{234} 
+ e^{i\omega}\sin\gamma|101\rangle_{234}), \label{eq61} \\
\hat{\sigma}^{(11)\dagger}|\text{W-like}\rangle_{234} =&& \frac{1}{\sqrt{2}}(-|001\rangle_{234} + e^{i\varphi}\cos\gamma|010\rangle_{234} 
+ e^{i\omega}\sin\gamma|100\rangle_{234}) . \label{eq62} 
\end{eqnarray}

Substituting eq.\eqref{eq63} into the right-hand side of eq.\eqref{eq58}, we confirm that the state $\{c_{pq}|\beta_{pq}\rangle\}$ form an orthonormal system:

\begin{equation}
{}_{123}\langle\beta_{mn}|c_{mn}^*c_{pq}|\beta_{pq}\rangle_{123}
=\frac{1}{4}\delta_{mp}\delta_{nq} . \label{eq64} 
\end{equation}

We find that the norm of each state $c_{pq}|\beta_{pq}\rangle$ is equal to 1/2: $||c_{pq}|\beta_{pq}\rangle|| = 1/2$ and the absolute value of each expansion coefficient $c_{pq}$ is equal to 1/2: $|c_{pq}|=1/2$. Therefore, if we ignore the phase of the expansion coefficients in eq.\eqref{eq57}, the expansion coefficients $\{c_{mn}\}$ and Alice's measurement basis $\{|\beta_{mn}\rangle\}$ are uniquely determined up to the 
unitary operator $\hat{S}$:

\begin{eqnarray}
c_{mn} &&= \frac{1}{2}, \label{eq65}\\
|\beta_{mn}\rangle_{123} &&= \sum_{\psi = 0}^1|\psi\rangle_1\otimes{}_4\langle\psi|\hat{U}^{(mn)\dagger}|\text{W-like}\rangle_{234} \nonumber\\ &&=\sum_{\psi = 0}^1|\psi\rangle_1\otimes{}_4\langle\psi|\hat{S}^{\dagger}\hat{\sigma}^{(mn)}|\text{W-like}\rangle_{234} ,  \label{eq66}
\end{eqnarray}

where in the last equation we used the definition eq.\eqref{eq50} of the unitary operator$\{\hat{U}^{(mn)}\}$. 
The result eq.\eqref{eq66} shows that the measurement consisting of the four bases performed by Alice contains the degrees of freedom coming from the unitary operator $\hat{S}$. In other words, Bob has one unitary transformation to be determined as he wishes. We can also check that $\hat{S} = \hat{I}$ holds for the four bases $\{|\beta_{mn}\rangle_{123}\}$ given by eq.\eqref{eq45}, eq.\eqref{eq46},  eq.\eqref{eq47}, and eq.\eqref{eq48}.

Finally, we compare the proof of impossibility of perfect quantum teleportation using the W state with the feasibility of one using the W-like state. The W state could not be used for perfect quantum teleportation because the  equality eq.\eqref{eq36} does not hold for the values of eq.\eqref{eq31} and eq.\eqref{eq32}. We can confirm that eq.\eqref{eq36} is satisfied when the W state is replaced by a W-like state.

We repeat the argument in Section \ref{sec:level3}, by replacing the W state with the W-like state. The reduced density matrix $\hat{\rho}_4$ in 
eq.\eqref{eq22} is replaced by:

\begin{eqnarray}
\hat{\rho}_4 =&& \operatorname{tr}_{123}(|\psi\rangle_{1}\langle\psi|_{1}\otimes|\text{W-like}\rangle_{234}\langle \text{W-like}|_{234}) \nonumber\\
=&&\operatorname{tr}_{1}(|\psi\rangle_{1}\langle\psi|_{1})\operatorname{tr}_{23}(|\text{W-like}\rangle_{234}\langle \text{W-like}|_{234}) \nonumber\\
=&&\operatorname{tr}_{23}(|\text{W-like}\rangle_{234}\langle \text{W-like}|_{234}) \nonumber\\
=&&\hat{I}/2 
=\frac{1}{2}|0\rangle_{4}\langle 0|_{4} +\frac{1}{2}|1\rangle_{4}\langle 1|_{4}\label{eq67} .
\end{eqnarray}

Therefore, eq.\eqref{eq30} is replaced by

\begin{eqnarray}
\frac{1}{2}|0\rangle_{4}\langle 0|_{4} +\frac{1}{2}|1\rangle_{4}\langle 1|_{4}=\sum^{1}_{p,q,r= 0}\hat{T}^{(pqr)}|0\rangle_{4}\langle0|_{4}\hat{T}^{(pqr)\dagger} \label{eq68}. \ 
\end{eqnarray}

By applying ${}_{4}\langle 0|\cdot|0\rangle_{4}$ and ${}_{4}\langle 1|\cdot|1\rangle_{4}$ to both sides of eq.\eqref{eq68} and using eq.\eqref{eq28}, eq.\eqref{eq31} and eq.\eqref{eq32} are replaced by

\begin{eqnarray}
\frac{1}{2} &&= \sum^{1}_{p,q,r= 0}|{}_4\langle0|\hat{T}^{(pqr)}|0\rangle_{4}|^2 =\sum^{1}_{p,q,r= 0}|{}_4\langle0|\hat{T}^{(pqr)}|1\rangle_{4}|^2  \label{eq69} , \ \\
\frac{1}{2} &&= \sum^{1}_{p,q,r= 0}|{}_4\langle1|\hat{T}^{(pqr)}|0\rangle_{4}|^2 =\sum^{1}_{p,q,r= 0}|{}_4\langle1|\hat{T}^{(pqr)}|1\rangle_{4}|^2  \label{eq70}. \ 
\end{eqnarray}

Using eq.\eqref{eq69} and eq.\eqref{eq70} shows that the equality \eqref{eq36} holds  for the W-like state although it was not satisfied for the W state.

Next, we consider what happens if the W-like state is replaced by the W state in the argument this section. In the W-like state, it is clear that the orthogonality of the measurement basis $\{|\beta_{pq}\rangle\}$ in eq.\eqref{eq66} results from the orthogonality of the states $\{\hat{\sigma}^{(pq)\dagger}|\text{W-like}\rangle_{234}\}$ in eq.\eqref{eq63}. It can be verified that states $\{\hat{\sigma}^{(pq)\dagger}|\text{W}\rangle_{234}\}$, which replaces the W-like state with the W state, is not orthonormal.
This fact can be checked by explicitly writing the states:

\begin{eqnarray}
\hat{\sigma}^{(00)\dagger}|\text{W}\rangle_{234} =&& \frac{1}{\sqrt{3}}(|001\rangle_{234} + |010\rangle_{234} 
+ |100\rangle_{234}), \label{eq71} \\
\hat{\sigma}^{(01)\dagger}|\text{W}\rangle_{234} =&& \frac{1}{\sqrt{3}}(|000\rangle_{234} + |011\rangle_{234} 
+ |101\rangle_{234}), \label{eq72} \\
\hat{\sigma}^{(10)\dagger}|\text{W}\rangle_{234} =&& \frac{i}{\sqrt{3}}(-|000\rangle_{234} + |011\rangle_{234} 
+ |101\rangle_{234}), \label{eq73} \\
\hat{\sigma}^{(11)\dagger}|\text{W}\rangle_{234} =&& \frac{1}{\sqrt{3}}(-|001\rangle_{234} +|010\rangle_{234} 
+ |100\rangle_{234}) . \label{eq74} 
\end{eqnarray}

For example, the value of the inner product ${}_4\langle\text{W}|\hat{\sigma}^{(11)}\hat{\sigma}^{(00)\dagger}|\text{W}\rangle_{234}$ is 
calculated from eq.\eqref{eq71} and eq.\eqref{eq74}: 

\begin{eqnarray}
{}_{234}\langle\text{W}|\hat{\sigma}^{(11)}\hat{\sigma}^{(00)\dagger}|\text{W}\rangle_{234} = \frac{1}{3} \neq 0.
\end{eqnarray}

Therefore, the measurement basis $\{|\beta_{pq}\rangle\}$ consisting of W states is not orthonormal. Then we have shown that the argument using the W-like state given in this section does not hold if the W-like state is replaced by the W state, which is compatible with the impossibility proof of perfect quantum teleportation using  the W state given in the previous section.

\section{\label{sec:level5}Alternative approach to perfect quantum teleportation}


\begin{figure}[h]
\centering
\begin{minipage}{0.49\columnwidth}
    \centering
    \includegraphics[width=0.9\columnwidth]{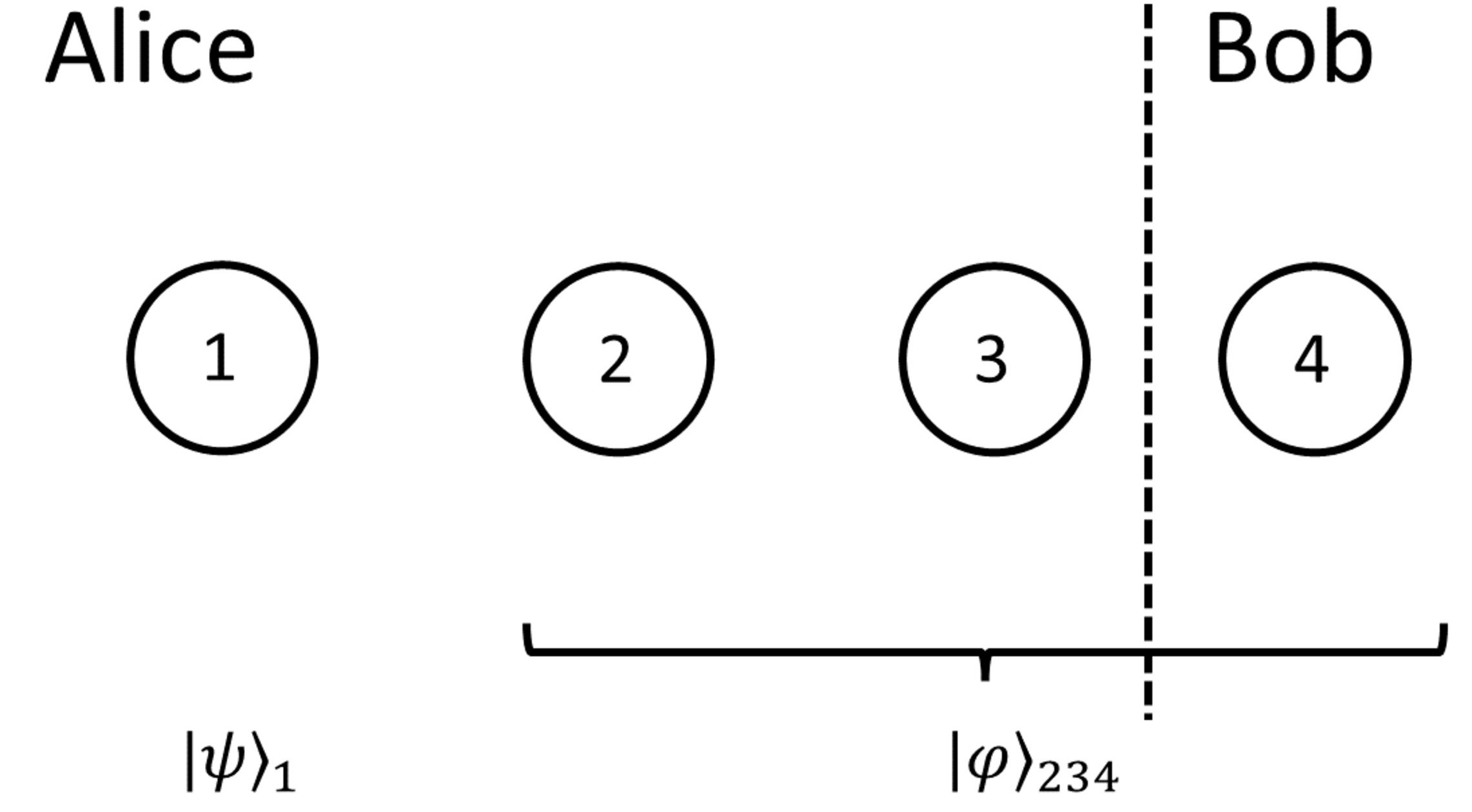}
    \caption{The initial state of Alice and Bob before the operation $\hat{U}_{23}$}
    \label{Figure1}
\end{minipage}
\hfill
\begin{minipage}{0.49\columnwidth}
    \centering
    \includegraphics[width=0.7\columnwidth]{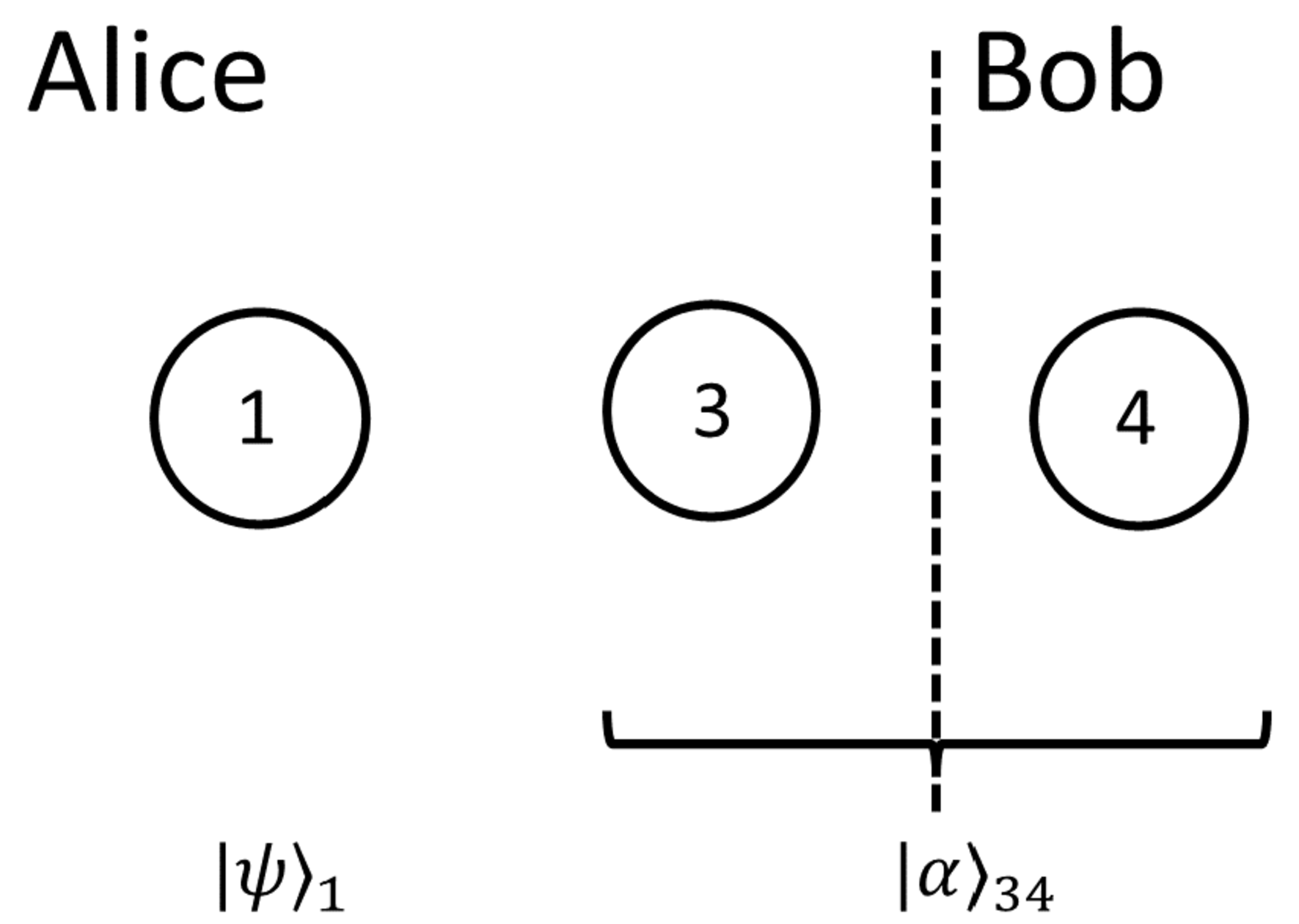}
    \caption{The initial state of Alice and Bob after the operation $\hat{U}_{23}$
    }
    \label{Figure2}
\end{minipage}
\end{figure}


Our setting of quantum teleportation for an unknown quantum state $|\psi\rangle_1$ involves sharing a 3-qubit state $|\varphi\rangle_{234}$ between Alice and Bob such that Alice holds two qubits (2 and 3) and Bob holds one qubit (4). (See Fig.\ref{Figure1}) 
In this configuration, genuine multipartite entanglement is not the key issue because the relevant factor is the bipartite entanglement between the partition (2,3) and 4. 
In view of this, it is important to recall the well-known fact \cite{R,S,WC,MH}: the necessary and sufficient condition for performing perfect quantum teleportation between two parties sharing a 2-qubit entangled state is that the shared state must be a maximally entangled state. 
If this fact can be applied also to sharing entangled states in a 3-qubit system in question, we have a chance of somewhat simplifying the lengthy algebraic proofs already given in the previous sections including the case of the W state.

Concretely, this alternative method is realized according to the following steps.
Suppose that there exists such a global unitary operator $\hat{U}_{23}$ which acts on the entangled state $|\varphi\rangle_{234}$ shared by Alice and Bob teleporting an unknown quantum state $|\psi\rangle_1$ (See Fig.\ref{Figure1}), such that
$\hat{U}_{23}$ transforms her two qubits into a tensor product $|0\rangle_2 \otimes |\alpha\rangle_{34}$ of $|0\rangle_2$ and some state $|\alpha\rangle_{234}$:
\begin{align}
\hat{U}_{23} \otimes \hat{I}_4|\varphi\rangle_{234} = |0\rangle_2 \otimes |\alpha\rangle_{34}. \label{eq*1}
\end{align}
The operator $\hat{U}_{23}$ is a local unitary transformation acting only on Alice's system 23 for the shared state $|\varphi\rangle_{234}$. Therefore, the entanglement between Alice (system 23) and Bob (system 4) remains unchanged. 
By separating the state $|0\rangle_2$ of system 2, we see that the shared state $|\alpha\rangle_{34}$ is the actual entangled state held by Alice and Bob. 

Now, the problem of whether or not Alice, who possesses an unknown quantum state $|\psi\rangle_1$, can perform perfect quantum teleportation to Bob's system 4 is equivalent to the problem of whether or not the shared state $|\alpha\rangle_{34}$ is a suitable entangled state for this purpose (See Fig.\ref{Figure2}). 
It is known that for the latter problem, it is both necessary and sufficient that the shared state $|\alpha\rangle_{34}$ is a maximally entangled state in order to perform complete quantum teleportation\cite{R,S,WC,MH} (indeed, the shared state $|\beta_{00}\rangle$ in eq.\eqref{eq1} is a maximally entangled state).
If this is the case, the remaining task is then to check whether the reduced 2-qubit entangled state between Alice and Bob is maximally entangled or not. 
If this is successful, by checking this process, it becomes clear without detailed calculations that the 3-qubit state$|\varphi\rangle_{234}$ in question is capable of perfect quantum teleportation.

In this section, we first explain the GHZ state as a successful example of this method (see Subsection \ref{sec:level5.1}). Next, we show that this method cannot be applied to the W state (see Subsection \ref{sec:level5.2}). Finally, we show that this method cannot be applied to  the W-like state except for a special case (see Subsection \ref{sec:level5.3}).

\subsection{\label{sec:level5.1}The case of the GHZ state}

We apply the alternative method to judge whether or not 
the GHZ state can be used as a shared state suitable for performing perfect quantum teleportation. 
In the GHZ state \eqref{eq6}, indeed, applying the global unitary operator defined by
\begin{align}
\hat{U}_{23} = |00\rangle_{23}\langle 00|_{23} +|01\rangle_{23}\langle 11|_{23} +|10\rangle_{23}\langle 01|_{23} +|11\rangle_{23}\langle 10|_{23} 
\end{align}
yields the relation:
\begin{align}
\hat{U}_{23} \otimes \hat{I}_4|\text{GHZ}\rangle_{234} = |0\rangle_2 \otimes \frac{1}{\sqrt{2}}(|00\rangle_{34} + |11\rangle_{34}),
\end{align}
which results in the form of eq.\eqref{eq*1}. In this case, the state $|\alpha\rangle_{34} = \frac{1}{\sqrt{2}}(|00\rangle_{34} + |11\rangle_{34})$ becomes the maximally entangled state. This observation demonstrates that the GHZ state can be used as a shared state for performing perfect quantum teleportation.

However, it will be shown in the next Subsection \label{sec:level5.2} that this method cannot be used to discuss the feasibility of perfect quantum teleportation using the W state. Specifically, it will be demonstrated that no global unitary operator $\hat{U}_{23}$ exists in the W state that satisfies the form specified by eq.\eqref{eq*1}.

\subsection{\label{sec:level5.2}The case of the W state}

We point out that there does not exist the global unitary operator $\hat{U}_{23}$ that satisfies eq.\eqref{eq*1} for the W state, in sharp contrast to the GHZ state. 
Indeed, this fact is demonstrated by showing that no unitary operator $\hat{U}_{23}$ simultaneously satisfies the following three relations: 
\begin{align}
\hat{U}_{23}|00\rangle_{23} =& |0\rangle_2 \otimes |\phi_a\rangle_3, \label{eq*2}\\
\hat{U}_{23}|01\rangle_{23} =& |0\rangle_2 \otimes |\phi_b\rangle_3, \label{eq*3}\\
\hat{U}_{23}|10\rangle_{23} =& |0\rangle_2 \otimes |\phi_c\rangle_3. \label{eq*4}
\end{align}
This is because the action of the global unitary operator $\hat{U}_{23}$ on the W state \eqref{eq15} given by 
\begin{align}
\hat{U}_{23}\otimes \hat{I}_4|\text{W}\rangle_{234} = \frac{1}{\sqrt{3}}(\hat{U}_{23}|00\rangle_{23}\otimes|1\rangle_4 + \hat{U}_{23}|01\rangle_{23}\otimes|0\rangle_4 +\hat{U}_{23}|10\rangle_{23}\otimes|0\rangle_4),
\end{align}
takes the form of equation \eqref{eq*1} only when the three relations eq.\eqref{eq*2}, eq.\eqref{eq*3}, and eq.\eqref{eq*4} are simultaneously satisfied.

Suppose that an operator $\hat{U}_{23}$ satisfying the three relations eq.\eqref{eq*2}, eq.\eqref{eq*3}, and eq.\eqref{eq*4} exists. Then, the state $\hat{U}_{23}|jk\rangle_{23}$ obtained by acting the operator $\hat{U}_{23}$ on the basis $|jk\rangle_{23}(j,k \in \{0,1\})$ is expanded with respect to the basis set $\{|\ell m\rangle_{23}\}$ as 
\begin{align}
\hat{U}_{23}|jk\rangle_{23} = \sum_{\ell ,m =0}^1u_{\ell m,jk}|\ell m\rangle_{23} \ , \ |\ell m\rangle_{23} = |\ell\rangle_2 \otimes |m\rangle_3 \ \ (u_{\ell m,jk} \in \mathbb{C}). \label{eq*4.5}
\end{align}
For the choices $|jk\rangle = |00\rangle,|01\rangle,|10\rangle$,  eq.\eqref{eq*4.5} is respectively written down:
\begin{align}
\hat{U}_{23}|00\rangle_{23} =& |0\rangle_2 \otimes (u_{00,00}|0\rangle_3 +u_{01,00}|1\rangle_3) \nonumber\\
&+|1\rangle_2\otimes (u_{10,00}|0\rangle_3 +u_{11,00}|1\rangle_3), \label{eq*5}\\
\hat{U}_{23}|01\rangle_{23} =& |0\rangle_2 \otimes (u_{00,01}|0\rangle_3 +u_{01,01}|1\rangle_3) \nonumber\\
&+|1\rangle_2\otimes (u_{10,01}|0\rangle_3 +u_{11,01}|1\rangle_3), \label{eq*6}\\
\hat{U}_{23}|10\rangle_{23} =& |0\rangle_2 \otimes (u_{00,10}|0\rangle_3 +u_{01,10}|1\rangle_3) \nonumber\\
&+|1\rangle_2\otimes (u_{10,10}|0\rangle_3 +u_{11,10}|1\rangle_3). \label{eq*7}
\end{align}
Then we find that the three equations eq.\eqref{eq*2}, eq.\eqref{eq*3}, and eq.\eqref{eq*4} hold when the conditions are satisfied:
\begin{align}
u_{10,00}=&u_{11,00}=u_{10,01}=u_{11,01}=u_{10,10}=u_{11,10}=0 \label{eq*8}, 
\end{align}
which implies
\begin{align}
|\phi_a\rangle_3 =& u_{00,00}|0\rangle_3 +u_{01,00}|1\rangle_3 \label{eq*9}, \\
|\phi_b\rangle_3 =& u_{00,01}|0\rangle_3 +u_{01,01}|1\rangle_3 \label{eq*10}, \\
|\phi_c\rangle_3 =& u_{00,10}|0\rangle_3 +u_{01,10}|1\rangle_3 \label{eq*11}. 
\end{align}
Since the matrix representation of the operator $\hat{U}_{23}$ in the basis $\{|\ell m\rangle_{23}\}$ is given by $u_{\ell m,jk} = {}_{23}\langle \ell m|\hat{U}_{23}|jk\rangle_{23}$:
\begin{align}
u:=
\begin{pmatrix}u_{00,00}&u_{00,01}&u_{00,10}&u_{00,11}\\
u_{01,00}&u_{01,01}&u_{01,10}&u_{01,11}\\
u_{10,00}&u_{10,01}&u_{10,10}&u_{10,11}\\
u_{11,00}&u_{11,01}&u_{11,10}&u_{11,11}\\
\end{pmatrix}, 
\end{align}
the conditions eq.\eqref{eq*8} lead to the form:
\begin{align}
u=
\begin{pmatrix}u_{00,00}&u_{00,01}&u_{00,10}&u_{00,11}\\
u_{01,00}&u_{01,01}&u_{01,10}&u_{01,11}\\
0&0&0&u_{10,11}\\
0&0&0&u_{11,11}\\
\end{pmatrix} 
\end{align}
Since the operator $\hat{U}_{23}$ is unitary, the representation matrix $u$ must be unitary. As properties to be satisfied by a unitary matrix, each row vector must have a unit norm, and distinct two row vectors must be orthogonal to each other. 
Therefore, both the third row vector $(0,0,0,u_{10,11})$ and  the fourth row vector $(0,0,0,u_{11,11})$ of the unitary matrix $u$ have norm 1, namely, $|u_{10,11}| = 1$ and $|u_{11,11}| = 1$, which are rewritten into $u_{10,11} = e^{ia}$ and $u_{11,11} = e^{ib}(i:=\sqrt{-1}, \ a,b\in \mathbb{R})$. As a result, the two vectors have the non-vanishing inner product $e^{i(a-b)} \neq 0$ because of $|e^{i(a-b)}| = 1$, and hence they never become orthogonal. Thus, we conclude that the operator $\hat{U}_{23}$ cannot be unitary.

Therefore, the determination of the feasibility of perfect quantum teleportation based on the global unitary operator, which was possible in the GHZ state, cannot be applied to the W state. Consequently, determining the feasibility of perfect quantum teleportation when the W state is used as the shared state requires the lengthy algebraic calculations described earlier in Section 3.

\subsection{\label{sec:level5.3}The case of the W-like state}

It has been shown in Section 4 that the W-like state \eqref{eq43} can be used for perfect quantum teleportation for any parameters $\varphi,\omega,\gamma \in \mathbb{R}$. Therefore, we investigate whether the method using the global unitary operator $\hat{U}_{23}$ that transforms the expression into the form of eq.\eqref{eq*1} is applicable or not.
Applying the global unitary operator $\hat{U}_{23}$ to the W-like state \eqref{eq43} yields 
\begin{align}
\hat{U}_{23}\otimes\hat{I}_4|\text{W-like}\rangle_{234} =&\frac{1}{\sqrt{2}}(\hat{U}_{23}|00\rangle_{23}\otimes|1\rangle_{4} + e^{i\varphi}\cos\gamma\hat{U}_{23}|01\rangle_{23}\otimes|0\rangle_{4} \nonumber\\
&+ e^{i\omega}\sin\gamma\hat{U}_{23}|10\rangle_{23}\otimes|0\rangle_{4}).
\end{align}

For values of parameter $\gamma$ satisfying $\cos \gamma \neq 0$ or $\sin \gamma \neq 0$ (e.g., the choice $\varphi=\omega =0,\gamma =\frac{\pi}{4}$ leads to $|\text{W-like}\rangle_{234} = \frac{1}{2}(\sqrt{2}|001\rangle_{234}+|010\rangle_{234}+|100\rangle_{234})$), we find that, similarly to the W state,
it is impossible to determine the feasibility of perfect quantum teleportation since there exists no global unitary operator $\hat{U}_{23}$ that simultaneously satisfies eq.\eqref{eq*2}, eq.\eqref{eq*3}, and eq.\eqref{eq*4}. 

However, for example,
in the W-like state \eqref{eq43} where $\cos \gamma = 0(\sin \gamma = 1)$, that is, when $\gamma = \frac{\pi}{2}$, adopting the global unitary operator $\hat{U}_{23}$ given by
\begin{align}
\hat{U}_{23} = |00\rangle_{23}\langle 00|_{23} + |01\rangle_{23}\langle 10|_{23} + |10\rangle_{23}\langle 01|_{23} + |11\rangle_{23}\langle 11|_{23} 
\end{align}
yields
\begin{align}
\hat{U}_{23}\otimes\hat{I}_4|\text{W-like}\rangle_{234}|_{\gamma = \frac{\pi}{2}} = |0\rangle_2\otimes\frac{1}{\sqrt{2}}(|01\rangle_{34} + e^{i\omega}|10\rangle_{34}),
\end{align}
which results in the form of eq.\eqref{eq*1}. In this case, the state $\frac{1}{\sqrt{2}}(|01\rangle_{34} + e^{i\omega}|10\rangle_{34})$ of system 34 is the maximally entangled state as expected, which enables the perfect quantum teleportation. 

\section{\label{sec:level6}Conclusion}

In this paper we have shown that the unknown quantum state of a 1-qubit system cannot be perfectly teleported by a basis measurement identifiable between two parties if the sharing state is the W state. The background to this is the state $\hat{\rho}_4$ owned by the receiver Bob among the two parties that shared the W state. 
From the form of eq.\eqref{eq22}, the state $\hat{\rho}_4$ is a mixture of orthonormal pure states $\{|0\rangle,|1\rangle\}$ in different proportions $\{2/3,1/3\}$. As a result, the proof of Section 3 shows that the equality of eq.\eqref{eq36} does not hold from the values of eq.\eqref{eq31} and eq.\eqref{eq32}. If Bob's state $\hat{\rho}_4$ is a maximal mixed state $(\hat{I}/2)$ in which orthonormal pure states are mixed in the same proportion $\{1/2,1/2\}$, then the equality in eq.\eqref{eq36} is satisfied. In fact, for perfect teleportation between two parties sharing the GHZ state, the state $\hat{\rho}_4$ of system 4 is suitable for teleportation as $\hat{I}/2$ from eq.\eqref{eq6}.

The proof of the impossibility of perfect quantum teleportation in the W state which was first given in Section 3 based on the algebraic method was naively expected to be simplified by an alternative method given by introducing a global unitary transformation $\hat{U}_{23}$ applied to Alice's system 23 constituting the shared state. However, it was confirmed that the alternative method is in general inapplicable except for specific entangled states.  In particular, it is not applicable to the W state as shown in Section \ref{sec:level5}.

The W-like state, which is obtained by modifying the coefficients of the W state, was proposed in response to this request, and has a structure in which the state owned by Bob is a maximal mixed state. 
We have shown that the W-like state is actually used for realizing the perfect quantum teleportation.
This result confirms the examples proposed in the preceding studies\cite{P,A} from a general viewpoint. 
If the unitary transformations appearing in Bob's system are determined in advance in the form of eq.\eqref{eq50}, then the expansion coefficients appearing in the quantum teleportation equation and the measurement basis of the sender Alice are found to be determined in the form of eq.\eqref{eq36} and eq.\eqref{eq66}. 
Here it should be remarked that one of the unitary transformations in eq.\eqref{eq50} can be set as any unitary transformation that Bob desires, and all the other unitary transformations can be realized by letting the identity operator and the Pauli operators act on the prepared operator. 
These results give a convenient way to set up the perfect quantum teleportation from the receiver Bob's point of view.

All entangled states discussed in this paper are three-qubit states composed of two-level systems. From the viewpoint of bipartite entanglement entropy, the GHZ state and the W-like states provide one full ebit of entanglement between the two parties and can therefore serve as shared resources for perfect quantum teleportation.
By contrast, the entanglement entropy of the W state, when adopted as the shared state, is strictly smaller than one ebit. As a result, it does not meet the necessary entanglement requirement for perfect deterministic quantum teleportation. This criterion thus also leads to the conclusion that the W state cannot be used as a suitable shared state for perfect quantum teleportation (see \ref{sec:Appendix}).
In the present setup, Bob’s subsystem is a qubit, so the Schmidt rank of the shared bipartite state is at most two; accordingly, the entanglement entropy uniquely determines whether one full ebit of entanglement is available.



\appendix

\section{\label{sec:Appendix}Entanglement entropy criterion for perfect teleportation in the present setup}


In this appendix, we examine the feasibility of perfect quantum teleportation from the viewpoint of bipartite entanglement between Alice and Bob using the entanglement entropy.

In the present setup, Alice holds subsystems 2 and 3, while Bob holds subsystem 4. The shared three-qubit state $|\varphi\rangle_{234}$ can therefore be regarded as a bipartite pure state $|\varphi\rangle_{AB}$, where $A=(2,3)$ and $B=4$.
Since Bob's subsystem $B$ is a two-level system, the Schmidt rank of the bipartite state $|\varphi\rangle_{AB}$ is at most two. In this situation, the Schmidt coefficients are uniquely determined (up to permutation) by the eigenvalues of Bob's reduced density matrix
\begin{equation}
\hat{\rho}_B = \mathrm{tr}_A\bigl(|\varphi\rangle_{AB}\langle\varphi|_{AB}\bigr).
\end{equation}
Consequently, the entanglement entropy
\begin{equation}
S(\hat{\rho}_B) = -\mathrm{tr}_B\bigl(\hat{\rho}_B \log_2 \hat{\rho}_B\bigr)
\end{equation}
provides a complete criterion for determining whether Alice and Bob share one full ebit of entanglement in the present teleportation setup. In what follows, we use this entanglement entropy to assess whether a given shared state is suitable for perfect quantum teleportation.


As shown in Sections~2 and~4, both the GHZ state and the W-like state can serve as shared resources for perfect quantum teleportation. 
Indeed, when either of these states is adopted as the shared state between Alice and Bob, Bob's reduced density operator
$\hat{\rho}_B$ becomes the maximally mixed state:
\begin{equation}
\hat{\rho}_B = \frac{\hat{I}_B}{2},
\end{equation}
as shown below:
\begin{align}
\hat{\rho}_B\Big|_{|\varphi\rangle_{AB} = |\text{GHZ}\rangle_{AB}}=&\operatorname{tr}_A(|\text{GHZ}\rangle_{AB}\langle\text{GHZ}|_{AB}) \nonumber\\
=&\operatorname{tr}_A\left\{\frac{1}{\sqrt{2}}(|000\rangle_{AB} + |111\rangle_{AB})\frac{1}{\sqrt{2}}(\langle000|_{AB} + \langle111|_{AB})\right\} \nonumber\\
=&\frac{1}{2}\operatorname{tr}_A(|00\rangle_{A}\langle00|_{A}\otimes|0\rangle_{B}\langle0|_{B} + |00\rangle_{A}\langle11|_{A}\otimes|0\rangle_{B}\langle1|_{B} \nonumber\\
&+ |11\rangle_{A}\langle00|_{A}\otimes|1\rangle_{B}\langle0|_{B} + |11\rangle_{A}\langle11|_{A}\otimes|1\rangle_{B}\langle1|_{B}) \nonumber\\
=&\frac{1}{2}(|0\rangle_{B}\langle0|_{B} + |1\rangle_{B}\langle1|_{B}) 
=\frac{\hat{I}_B}{2},
\\
\hat{\rho}_B\Big|_{|\varphi\rangle_{AB} = |\text{W-like}\rangle_{AB}}=&\operatorname{tr}_A(|\text{W-like}\rangle_{AB}\langle\text{W-like}|_{AB}) \nonumber\\ 
=&\operatorname{tr}_A\left\{\frac{1}{\sqrt{2}}(|001\rangle_{AB} + e^{i\varphi}\cos\gamma|010\rangle_{AB} + e^{i\omega}\sin\gamma|100\rangle_{AB})\right.\nonumber\\
&\left.\frac{1}{\sqrt{2}}(\langle001|_{AB} +e^{-i\varphi}\cos\gamma \langle010|_{AB} + e^{-i\omega}\sin\gamma\langle100|_{AB})\right\} \nonumber\\
=&\frac{1}{2}\operatorname{tr}_A(|00\rangle_{A}\langle00|_{A}\otimes|1\rangle_{B}\langle1|_{B} \nonumber\\
&+ |00\rangle_{A}\langle01|_{A}\otimes e^{-i\varphi}\cos\gamma|1\rangle_{B}\langle0|_{B} \nonumber\\
&+ |00\rangle_{A}\langle10|_{A}\otimes e^{-i\omega}\sin\gamma|1\rangle_{B}\langle0|_{B} \nonumber\\
&+ |01\rangle_{A}\langle00|_{A}\otimes e^{i\varphi}\cos\gamma|0\rangle_{B}\langle1|_{B} \nonumber\\
&+ |01\rangle_{A}\langle01|_{A}\otimes \cos^2\gamma|0\rangle_{B}\langle0|_{B} \nonumber\\
&+ |01\rangle_{A}\langle10|_{A}\otimes e^{i(\varphi-\omega)}\cos\gamma\sin\gamma|0\rangle_{B}\langle0|_{B} \nonumber\\
&+ |10\rangle_{A}\langle00|_{A}\otimes e^{i\omega}\sin\gamma|0\rangle_{B}\langle1|_{B} \nonumber\\
&+ |10\rangle_{A}\langle01|_{A}\otimes e^{i(\omega-\varphi)}\cos\gamma\sin\gamma|0\rangle_{B}\langle0|_{B} \nonumber\\
&+ |10\rangle_{A}\langle10|_{A}\otimes \sin^2\gamma|0\rangle_{B}\langle0|_{B})\nonumber\\
=&\frac{1}{2}((\cos^2\gamma+\sin^2\gamma)|0\rangle_{B}\langle0|_{B} + |1\rangle_{B}\langle1|_{B}) 
=\frac{\hat{I}_B}{2}.
\end{align}
As a result, the entanglement entropy satisfies
\begin{align}
S(\hat{\rho}_B) = -\operatorname{tr}_B(\hat{\rho}_B\log_2\hat{\rho}_B) 
=-\operatorname{tr}_B\left(\frac{\hat{I}_B}{2}\log_2\frac{\hat{I}_B}{2}\right) 
=-\sum_{0,1}1\frac{1}{2}\log_2\frac{1}{2} 
= 1. 
\end{align}
This implies that Alice and Bob share one full ebit of entanglement, which is sufficient for perfect deterministic quantum teleportation.
In other words, both the GHZ state and the W-like state are two-party maximal entangled states.

By contrast, when the W state is used as the shared state, Bob's reduced density operator is given by
\begin{align}
\hat{\rho}_B=&\operatorname{tr}_A(|\text{W}\rangle_{AB}\langle\text{W}|_{AB}) \nonumber\\ 
=&\operatorname{tr}_A\left\{\frac{1}{\sqrt{3}}(|001\rangle_{AB} + |010\rangle_{AB} + |100\rangle_{AB})
\frac{1}{\sqrt{3}}(\langle001|_{AB} +\langle010|_{AB} + \langle100|_{AB})\right\} \nonumber\\
=&\frac{1}{3}\operatorname{tr}_A(|00\rangle_{A}\langle00|_{A}\otimes|1\rangle_{B}\langle1|_{B} + |00\rangle_{A}\langle01|_{A}\otimes|1\rangle_{B}\langle0|_{B} \nonumber\\
&+ |00\rangle_{A}\langle10|_{A}\otimes |1\rangle_{B}\langle0|_{B} + |01\rangle_{A}\langle00|_{A}\otimes |0\rangle_{B}\langle1|_{B}) \nonumber\\
&+ |01\rangle_{A}\langle01|_{A}\otimes |0\rangle_{B}\langle0|_{B} + |01\rangle_{A}\langle10|_{A}\otimes |0\rangle_{B}\langle0|_{B}) \nonumber\\
&+ |10\rangle_{A}\langle00|_{A}\otimes |0\rangle_{B}\langle1|_{B} + |10\rangle_{A}\langle01|_{A}\otimes |0\rangle_{B}\langle0|_{B}) \nonumber\\
&+ |10\rangle_{A}\langle10|_{A}\otimes |0\rangle_{B}\langle0|_{B})\nonumber\\
=&\frac{2}{3}|0\rangle_{B}\langle0|_{B} + \frac{1}{3}|1\rangle_{B}\langle1|_{B}, 
\end{align}
leading to the entanglement entropy $S(\hat{\rho}_B)$ of the W state with strictly smaller than one ebit:
\begin{align}
S(\hat{\rho}_B) =& -\operatorname{tr}_B(\hat{\rho}_B\log_2\hat{\rho}_B) \nonumber\\
=&-\operatorname{tr}_B\left\{\left(\frac{2}{3}|0\rangle_B\langle 0|_B + \frac{1}{3}|1\rangle_B\langle 1|_B\right)\left(\log_2\frac{2}{3}|0\rangle_B\langle 0|_B + \log_2\frac{1}{3}|1\rangle_B\langle 1|_B\right)\right\} \nonumber\\
=&-\frac{2}{3}\log_2\frac{2}{3} - \frac{1}{3}\log_2\frac{1}{3} 
= 0.9182\cdots < 1.
\end{align}
Therefore, Alice and Bob do not share one full ebit of entanglement when the W state is adopted as the shared state.
This means that the W state is not the maximally entangled state between two parties.

This demonstrates that the W state cannot serve as a resource for perfect deterministic quantum teleportation \textit{in the present setup}.
We emphasize that, although entanglement entropy alone does not generally classify bipartite pure states up to local unitary equivalence, \textit{it provides a necessary and sufficient criterion for assessing the availability of one full ebit of entanglement in the specific case considered here, where Bob's subsystem is a qubit}.


\section*{Acknowledgments}

The authors would like to express sincere thanks to the anonymous referee for giving valuable comments which were greatly helpful in improving the manuscript. 

S.K. was supported by JST SPRING, Grant Number JPMJSP2109.  
K.-I. K. was supported by Grant-in-Aid for Scientific Research, JSPS KAKENHI Grant No. (C) 23K03406.

\end{document}